\documentclass[acs,aelm,showkeys,preprint,amsmath,amssymb]{revtex4-1}

\usepackage[sort&compress]{natbib}
\usepackage{graphicx}
\usepackage{epsfig}
\usepackage{epstopdf}
\usepackage{subfigure}
\usepackage{float}
\usepackage{ifpdf}
\usepackage{bm}
\ifpdf
\usepackage[pdftex,
        colorlinks=true,
        pdftitle={},
        pdfpagemode=UseNone,
        bookmarksopen=true
        pdfoutput=1
        ]{hyperref}
\fi

\begin{document}

\title{Doping Dependence of the Second Magnetization Peak, Critical Current Density and Pinning Mechanism in BaFe$_{2-x}$Ni$_x$As$_2$ Pnictide Superconductors}
\author{Shyam Sundar$^{\ast,\dagger}$, Said Salem-Sugui Jr.$^{\dagger}$, Edmund Lovell$^{\ddagger}$, Alexander Vanstone$^{\ddagger}$, Lesley F Cohen$^{\ddagger}$, Dongliang Gong$^{\mathparagraph}$, Rui Zhang$^{\mathsection}$, Xingye Lu$^{\amalg}$, Huiqian Luo$^{\mathparagraph}$, Luis Ghivelder$^{\dagger}$.}

\affiliation{$^{\dagger}$Instituto de Fisica, Universidade Federal do Rio de Janeiro, 21941-972 Rio de Janeiro, RJ, Brazil}
\affiliation{$^{\ddagger}$The Blackett Laboratory, Physics Department, Imperial College London, London SW7 2AZ, United Kingdom}
\affiliation{$^{\mathparagraph}$Beijing National Laboratory for Condensed Matter Physics, Institute of Physics, Chinese Academy of Sciences, Beijing 100190, People's Republic of China}
\affiliation{$^{\mathsection}$Department of Physics and Astronomy, Rice University, Houston, Texas 77005, USA}
\affiliation{$^{\amalg}$Center for Advanced Quantum Studies and Department of Physics, Beijing Normal University, Beijing 100875, China}
\email{shyam.phy@gmail.com}

\begin{abstract}

A series of high quality BaFe$_{2-x}$Ni$_x$As$_2$ pnictide superconductors were studied using magnetic relaxation and isothermal magnetic measurements in order to study the second magnetization peak (SMP) and critical current behaviour in Ni-doped 122 family. The temperature dependence of the magnetic relaxation rate suggests a pinning crossover, whereas, it's magnetic field dependence hints a vortex-lattice structural phase-transition. The activation energy ($U$) estimated using the magnetic relaxation data was analyzed in detail for slightly-underdoped, slightly-overdoped and an overdoped samples, using Maley's method and collective creep theory. Our results confirm that the SMP in these samples is due to the collective (elastic) to plastic creep crossover as has been observed for the other members of 122-family. In addition, we also investigated the doping dependence of the critical current density ($J_c$) and the vortex-pinning behaviour in these compounds. The observed $J_c$ is higher than the threshold limit (10$^5$ A/cm$^2$) considered for the technological potential and even greater than 1 MA/cm$^2$ for slightly underdoped Ni-content, x = 0.092 sample. The pinning characteristics were analyzed in terms of the models developed by Dew-Hughes and Griessen $et$ $al$, which suggest the dominant role of $\delta l$-type pinning. 

\end{abstract}


\maketitle

\section{Introduction}

The study of vortex dynamics in type-II superconductors gained the interest of experimentalists and theoreticians as soon as the creep phenomenon in magnetization was observed in conventional low $T_c$ systems \cite{kim62, and62, bea69}. From a technological point of view, the creep behaviour in magnetization is directly related to a creep in the critical current showing the importance to understand the vortex pinning mechanism. Later, the study of vortex dynamics gained attention in the late 80s with the discovery of the high-$T_c$ cuprates, which shows an intrinsic giant thermal activated magnetic relaxation \cite{yes88} as well as the so called second magnetization peak (SMP) effect in the isothermal magnetization curves which renders a peak in the critical currents \cite{abu96, ros05, bar10, ros07}, as also observed in the low $T_c$ superconductors, such as, Nb \cite{sta04}. More recently (2008), the study of vortex dynamics regained the attention of the scientific community due to the discovery of the iron-pnictide and iron selenide superconductors \cite{kam08, hsu08, hai08, zhi08} with a moderately high-$T_c$ (from 20 K up to 56 K) \cite{ste11}, large upper critical fields, $H_{c2}$ \cite{sen08, jar08}, small anisotropy \cite{wang15, yua09, alt08} and better intergrain connectivity than the cuprates \cite{kat11, dur11}. These salient features of iron pnictide superconductors are potentially suitable for applications purposes \cite{hos18}. Besides, pnictides are known as multiband superconductors, which may play a role on the pinning of vortices through the inter-band and intra-band electron scatterings \cite{thu82}. Since then, vortex dynamics studies were performed on different pnictides compounds discovered over the years \cite{zhi08, pro08, kop10, li11, miu12, per13, su14, ss17a, yon18}, and, most of them are devoted to the study of the mechanism responsible for the appearance of SMP in isothermal magnetization curves. Contrary to the cuprates, where the SMP is mostly observed only for $H$ $\parallel$ c-axis, in Fe-pnictides, due to the low anisotropy, it is observed for both, $H$ $\parallel$ c-axis and $H$ $\parallel$ ab planes. A rich variety of mechanisms were proposed as responsible for the SMP in different iron-pnictide superconductors, such as, crossover from elastic to plastic \cite{ss17a, ss17b, wei06}, order-disorder transition \cite{miu12, hec14} and vortex-lattice phase transitions \cite{kop10, pra11}. However, the mechanism responsible for SMP in Ni-doped BaFe$_2$As$_2$ pnictide superconductors is as yet unresolved \cite{said11, said13, su14}. Interestingly, in iron-pnictide superconductors, it has been observed that the existence of the SMP is doping dependent \cite{ahm17, yon18}. 

This motivated us to investigate the vortex-dynamics in a series of high quality BaFe$_{2-x}$Ni$_x$As$_2$ (x = 0.092, 0.108, 0.12, 0.15, 0.18, 0.065) pnictide superconductors. In addition to the detailed study of the SMP in different Ni-content samples, a complementary study of the critical current density and the pinning behaviour is also performed on all samples using magnetic relaxation and isothermal magnetic measurements. A detailed analysis of the magnetic relaxation data using Maley's method \cite{mal90} and collective pinning theory \cite{fei89} unambiguously shows that the SMP in Ni-doped BaFe$_2$As$_2$ compounds is due to the collective (elastic) to plastic creep crossover, which might be accompanied by a vortex-lattice structural phase transition, similar to the Co-doped BaFe$_2$As$_2$ superconductor. The critical current density is found to be higher than the threshold limit ($>$ 10$^5$ A/cm$^2$) considered for technological applications. The doping dependence of critical current density, $J_c(x)$, does not follow the variation of superconducting transition temperature with Ni-content, $T_c$(x), and shows a spike-feature at x = 0.092. The dominant pinning in these crystals is found to be related to the variation of the charge carrier mean free path, generally known as $\delta l$-type pinning. 

\section{Experimental Details}

\begin{figure}
\centering
\includegraphics[height=12cm]{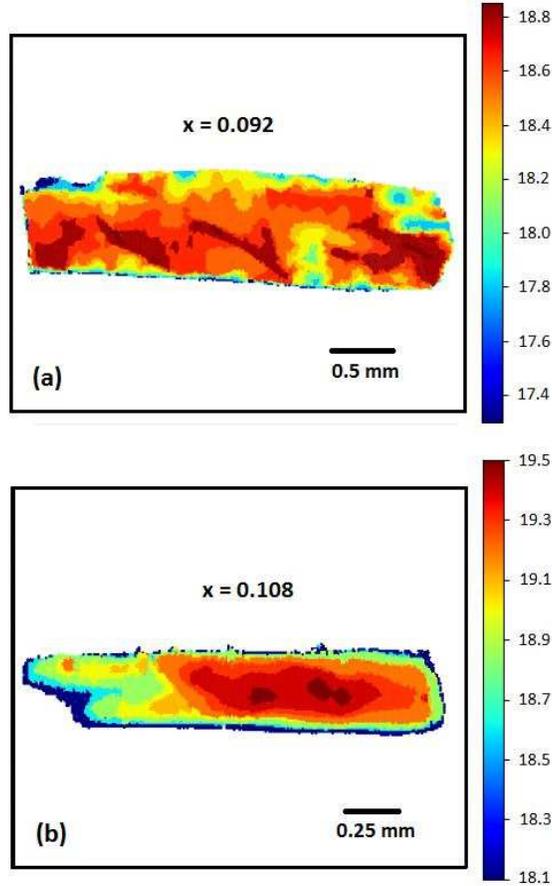}
\caption{\label{fig:Tcmap} Distribution of the superconducting transition temperature ($T_c$) in (a) x = 0.092 and (b) x = 0.108 samples, measured using a scanning Hall probe magnetometer. Variation of the $T_c$ over the scanned surface is identified by  labels in each panel. Both images show the good quality of the samples.}
\end{figure}

A detailed study on a series of six BaFe$_{2-x}$Ni$_x$As$_2$ pnictide superconductors is performed. Details of the crystal growth are described in Ref. \cite{yan11}. Large crystals were cut into small pieces with typical dimensions of 2.5 mm $\times$ 1 mm $\times$ 0.15 mm, using a clean scalpel and the samples with sharpest superconducting transition were chosen for each concentration to study. Surface maps of $T_c$, measured for two chosen samples (x = 0.092, 0.108) using a scanning Hall probe magnetometer with a 5 $\mu$m $\times$ $\mu$m active area of the Hall sensor (1 $\mu$m thick InSb epilayer on undoped GaAs substrate) \cite{per02} are shown in Fig. \ref{fig:Tcmap}. A 4 T split coil superconducting magnet and a continuous flow helium cryostat (Oxford Instruments Ltd.) were used to perform the measurements. The imaging was performed by applying 1 mT magnetic field (parallel to the c-axis) in zero field cooled state below $T_c$, and mapping the Meissner current profile across the crystal. At low temperature the screen current perfectly follows the edge of the sample. The mapping shows that the $T_c$ distribution within the crystals studied, are rather uniform and of high quality. The screening diminishes from the edges towards the center of the sample as expected within the measured transition width, consistent with the global magnetometry $M(T)$ data. Magnetization measurements were performed using a vibrating sample magnetometer (VSM, Quantum Design, USA), where the sample was mounted between the two quartz cylinders in a brass sample holder. The temperature and magnetic field dependence of the magnetization, $M(T)$, $M(H)$, and the magnetic relaxations, $M(t)$, were measured for $H$ $\parallel$ c-axis down to 2K and up to 9T magnetic field, in zero field cooled (zfc) mode. To investigate the behaviour of SMP in the samples, each $M(t)$ was measured over a period of approximately 90 minutes at fixed magnetic field in the increasing cycle of the isothermal, $M(H)$ curves. 

\section{Results and Discussion}

\begin{figure}
\centering
\includegraphics[height=10cm]{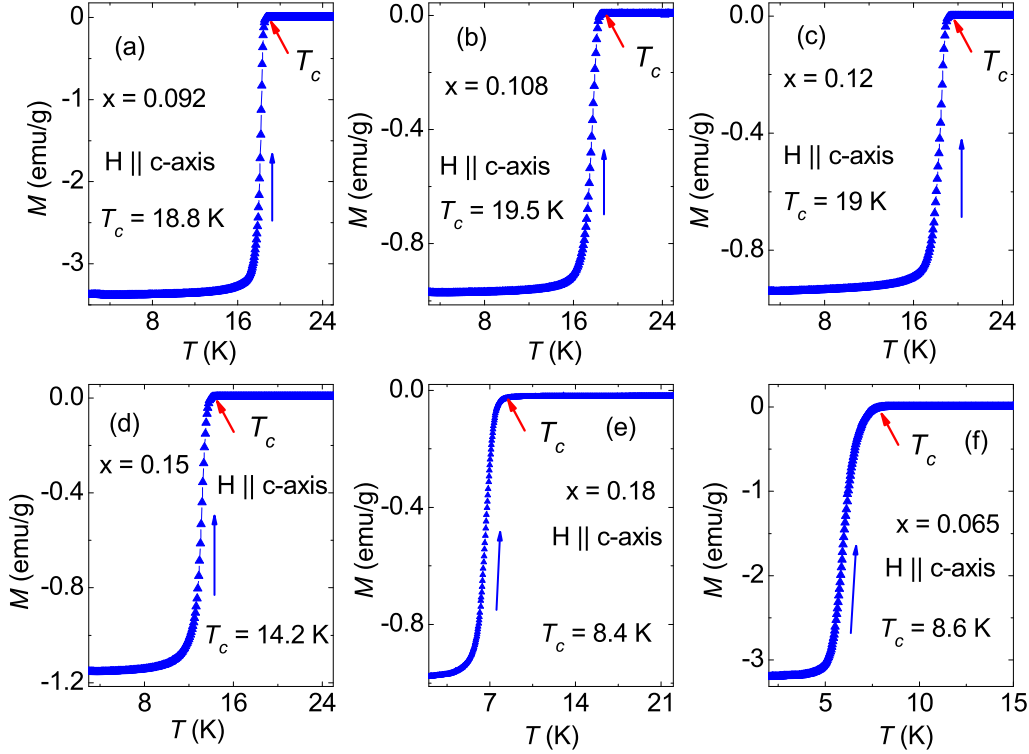}
\caption{\label{fig:Tc_All} (a-f) Temperature dependence of magnetization, $M(T)$, of BaFe$_{2-x}$Ni$_x$As$_2$ pnictide superconductors measured in zfc mode with $H$ = 1 mT.}
\end{figure}

Figure \ref{fig:Tc_All} shows the temperature dependence of magnetization, $M(T)$, measured for all samples in zfc mode with $H$ = 1 mT. The sharp drop in the magnetization for diamagnetic signal is considered as the onset of the superconducting transition ($T_c$), shown with an arrow in Fig. \ref{fig:Tc_All}. The sharp superconducting transition is an indication of the good quality of the samples and the obtained $T_c$ values are in fair agreement with the available literature \cite{yan11, wei16}. For BaFe$_{2-x}$Ni$_x$As$_2$ superconductors, the optimal doping is x = 0.1, with $T_c$ = 20.1 K \cite{yan11, wei16}. In the present study, the maximum $T_c$ = 19.5 K, is observed for x = 0.108, which is slightly overdoped and for more overdoped samples $T_c$ decreases. Similarly, x = 0.092 is slightly underdoped and shows the $T_c$ = 18.8 K, which further decreases for more underdoped samples. 

\subsection{Magnetic relaxation and the second magnetization peak (SMP)} 

\begin{figure}
\centering
\includegraphics[height=8cm]{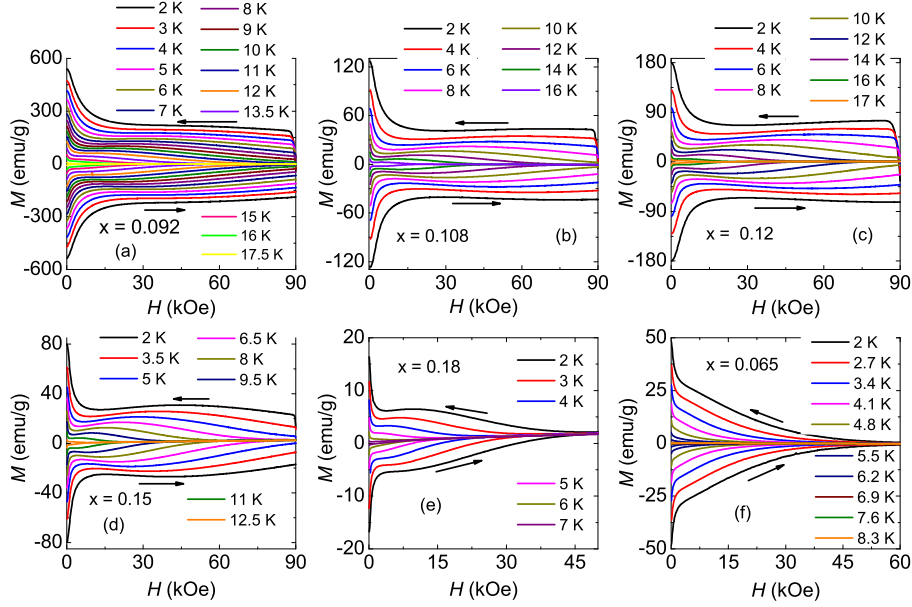}
\caption{\label{fig:MH_All} Isothermal magnetic field dependence of magnetization, $M(H)$, for x = (a) 0.092, (b) 0.108, (c) 0.12, (d) 0.15, (e) 0.18, and (f) 0.065 BaFe$_{2-x}$Ni$_x$As$_2$ pnictide superconductors, in field increasing and field decreasing cycle. Each sample show the second magnetization peak  feature below $T_c$, except the highly underdoped, x = 0.065 sample.}
\end{figure}

\begin{figure}
\centering
\includegraphics[height=8cm]{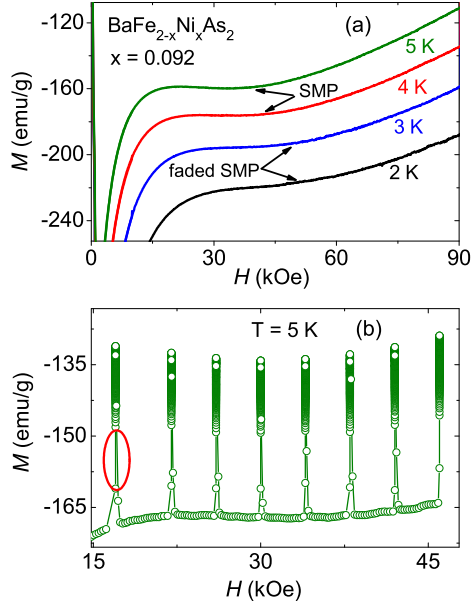}       
\caption{\label{fig:SMP_0.092} (a) Isothermal $M(H)$ for BaFe$_{2-x}$Ni$_x$As$_2$, x = 0.092 sample at some selected temperatures, where, below T = 4 K, the SMP feature smeared out. (b) Isothermal $M(H)$ at $T$ = 5 K with magnetic relaxation data measured for selected magnetic fields. The circle highlights the rapid magnetic relaxations for first 15 seconds.}
\end{figure}

Figure \ref{fig:MH_All} shows selected isothermal $M(H)$ curves, measured in zfc mode at various temperatures below $T_c$ down to 2 K. The symmetric behaviour of isothermal $M(H)$ suggests the dominant role of bulk pinning for all samples under study. A clear signature of the SMP is observed for each doping content, except for the highly underdoped, x = 0.065 sample. The absence of SMP in x = 0.065 sample is might be due to the static antiferromagnetic long-range order, which exists in low doped BaFe$_{2-x}$Ni$_x$As$_2$ samples \cite{wan10}. The onset and the peak position of the SMP are defined as $H_{on}$ and $H_P$ respectively. The magnetic hysteresis in the field increasing and field decreasing cycles of the $M(H)$ vanishes at higher fields, defined as the irreversibility field, $H_{irr}$. Interestingly, in slightly underdoped composition, x = 0.092, the SMP is smeared out below 5 K. This is called faded SMP, as seen in Fig. \ref{fig:SMP_0.092} (a). To confirm this anomaly, we repeated the measurements on another crystal with same x content (same $T_c$) and observed the same behaviour. Similar anamolous behaviour has also been observed in Bi-Sr-Ca-Cu-O single crystal, where, the SMP was only observed in a temperature range of 20-40 K below $T_c$ \cite{yes94, tam93}. In contrast to that, a recent study on Ba$_{0.75}$K$_{0.25}$Fe$_2$As$_2$ superconductor showed the SMP only at temperatures below $T_c$/2 and vanished at higher temperatures \cite{ss17a}.

To investigate the origin of the SMP in BaFe$_{2-x}$Ni$_x$As$_2$, with x = 0.092 (slightly underdoped), x = 0.108 (slightly overdoped), x = 0.15 (overdoped) superconductors, we performed magnetic relaxation, $M(t)$ at selected temperature and magnetic field values for $\sim$ 90 minutes in the lower branch of the $M(H)$ curves. In Fig. \ref{fig:SMP_0.092} (b), magnetic relaxation results are shown for x = 0.092 at $T$ = 5 K. A circle in Fig. \ref{fig:SMP_0.092} (b) highlights the initial 15 seconds of relaxation, which corresponds to $\sim$ 40 \% of the total magnetic relaxation in a period of 90 minutes of measurement. This feature is observed in all samples under investigation and is also found in a recent study on Ba$_{0.75}$K$_{0.25}$Fe$_2$As$_2$ \cite{ss17a}. All magnetic relaxations follow the usual logarithmic behaviour with time, $\mid$$M$$\mid$ $\sim$ log($t$) and the plots of ln$\mid$$M$$\mid$ vs ln$t$ allowed us to obtain the relaxation rate, $R$ = dln$\mid$$M$$\mid$/dln$t$.  

\begin{figure}
\centering
\includegraphics[height=8cm]{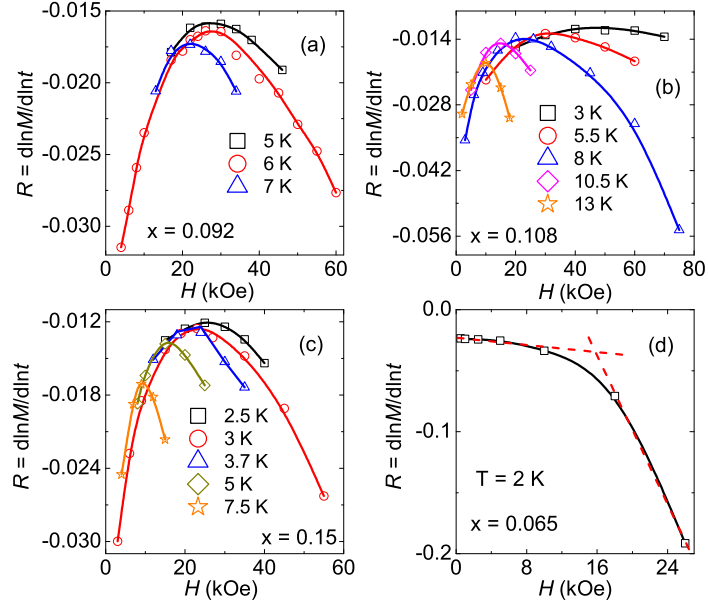}     
\caption{\label{fig:RH_All} Magnetic field dependence of relaxation rate, $R$ = dln$\mid$$M$$\mid$/dln$t$, for BaFe$_{2-x}$Ni$_x$As$_2$, x = (a) 0.092, (b) 0.108, (c) 0.15 and (d) 0.065 samples. Each sample shows the clear peak structure in every isothermal $R(H)$, except x = 0.065 sample, which also does not show the second magnetization peak feature.}
\end{figure}

Figure \ref{fig:RH_All} shows the relaxation rate as a function of magnetic field for samples with x = 0.092, 0.108, 0.15 and 0.065. For each Ni content, a peak in $R(H)$ associated to the SMP is observed in each curve. A similar feature has also been observed in the SMP study of Co-doped BaFe$_2$As$_2$ and explained in terms of the vortex-lattice structural phase transition \cite{ss17b, kop10}. It is worth mentioning the absence of peak in $R(H)$ for x = 0.065 sample, which do not show the SMP.

\begin{figure}
\centering
\includegraphics[height=10cm]{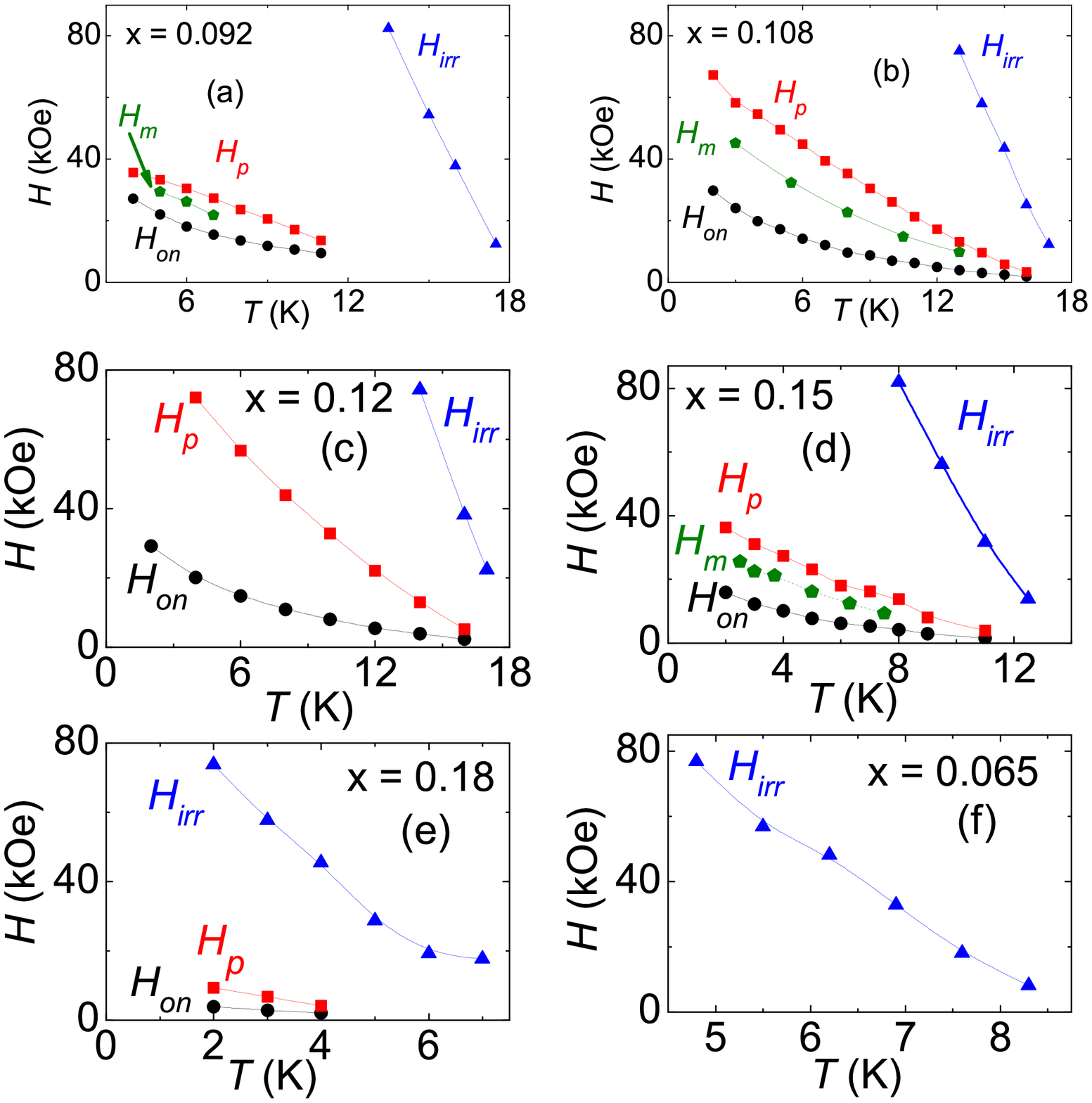}       
\caption{\label{fig:HT_diagram} The $H$-$T$ phase-diagram for x = (a) 0.092, (b) 0.108, (c) 0.12, (d) 0.15, (e) 0.18, and (f) 0.065 samples. The characteristic fields, $H_{on}$, $H_p$, $H_{irr}$, and $H_m$ are well explained in the text. }
\end{figure}

The characteristic magnetic fields associated with the SMP, $H_{on}$ and $H_p$, $H_{irr}$, and $H_m$ are shown in Fig. \ref{fig:HT_diagram} with $H_{on}$ and $H_p$ lying far below the $H_{irr}$ line. It should be noted that the behaviour of the temperature dependence of $H_p$ is different in x = 0.092, compared to the other samples used in this study. It should also be noted that $H_m$ lies in between the $H_{on}$ and $H_p$ lines as previously observed in the case of Co-doped BaFe$_2$As$_2$ \cite{ss17b, kop10}. Since, the peak position, $H_m$, in $H$-$T$ phase diagram varies with temperature in a similar way as observed for Ba(Fe$_{0.925}$Co$_{0.075}$)$_2$As$_2$ \cite{kop10}, we suggest that this behaviour might be associated with the vortex-lattice structural phase transition in the present study. However, it is argued that such vortex-lattice structural phase transition may be followed by a crossover in creep behaviour \cite{kop10, ss17b, ros05}. 

\begin{figure}
\centering
\includegraphics[height=8cm]{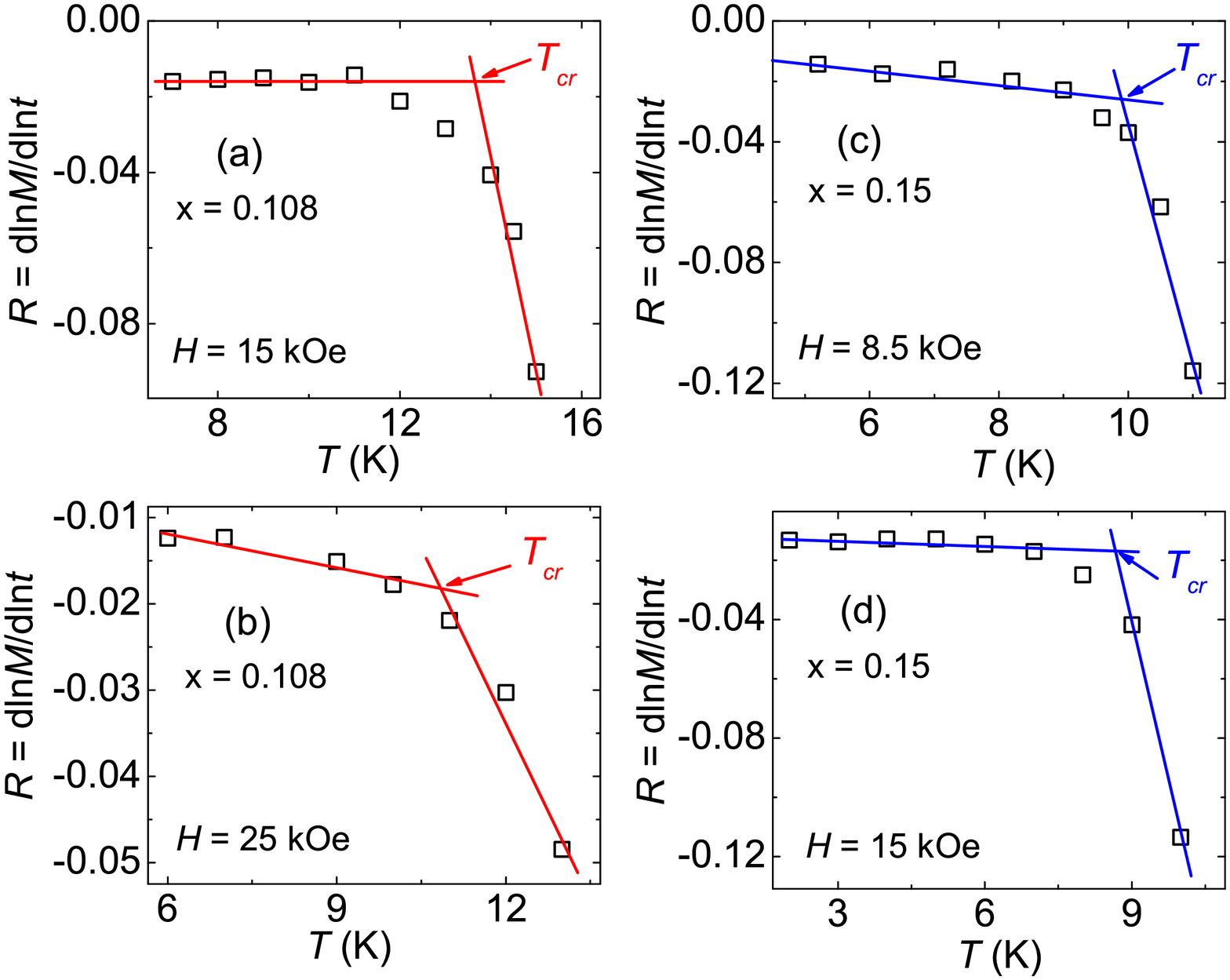} 
\caption{\label{fig:RT_All} Temperature dependence of relaxation rate, $R$ = dln$\mid$$M$$\mid$/dln$t$, for BaFe$_{2-x}$Ni$_x$As$_2$, x = (a, b) 0.108 and (c, d) 0.15 samples. Each isofield $R(T)$ shows a crossover in slope, which is defined as $T_{cr}$. The crossover in the slope suggests the crossover in pinning behaviour and is apparently related to the second magnetization peak ($H_p$) in the sample.}
\end{figure}

Figure \ref{fig:RT_All}, shows the temperature dependence of the relaxation rate, $R(T)$, for x = 0.108 and 0.15 measured with different magnetic fields. Each isofied $R(T)$, for both samples, shows a clear change of slope at $T_{cr}$. Interestingly, $T_{cr}$ values obtained from Fig. \ref{fig:RT_All} (a)-(d) are well matched with the $H_p$ line in the $H$-$T$ phase-diagram suggesting that a pinning crossover is responsible for the SMP. A peak behaviour observed simultaneously in isofield $R(T)$ and $R(H)$ has been argued as a possibility for a vortex-lattice structural phase-transition in different superconductors \cite{kop10, pra11, bro04}, but in the present case the peak positions of R(H) and R(T) do not match. As we see in Fig. \ref{fig:RH_All}, each isothermal $R(H)$ shows a peak behaviour, however, the isofield $R(T)$ shown in Fig. \ref{fig:RT_All} (a-d), only show a change of slope and do not exhibit a clear peak structure.
\begin{figure}
\centering
\includegraphics[height=8cm]{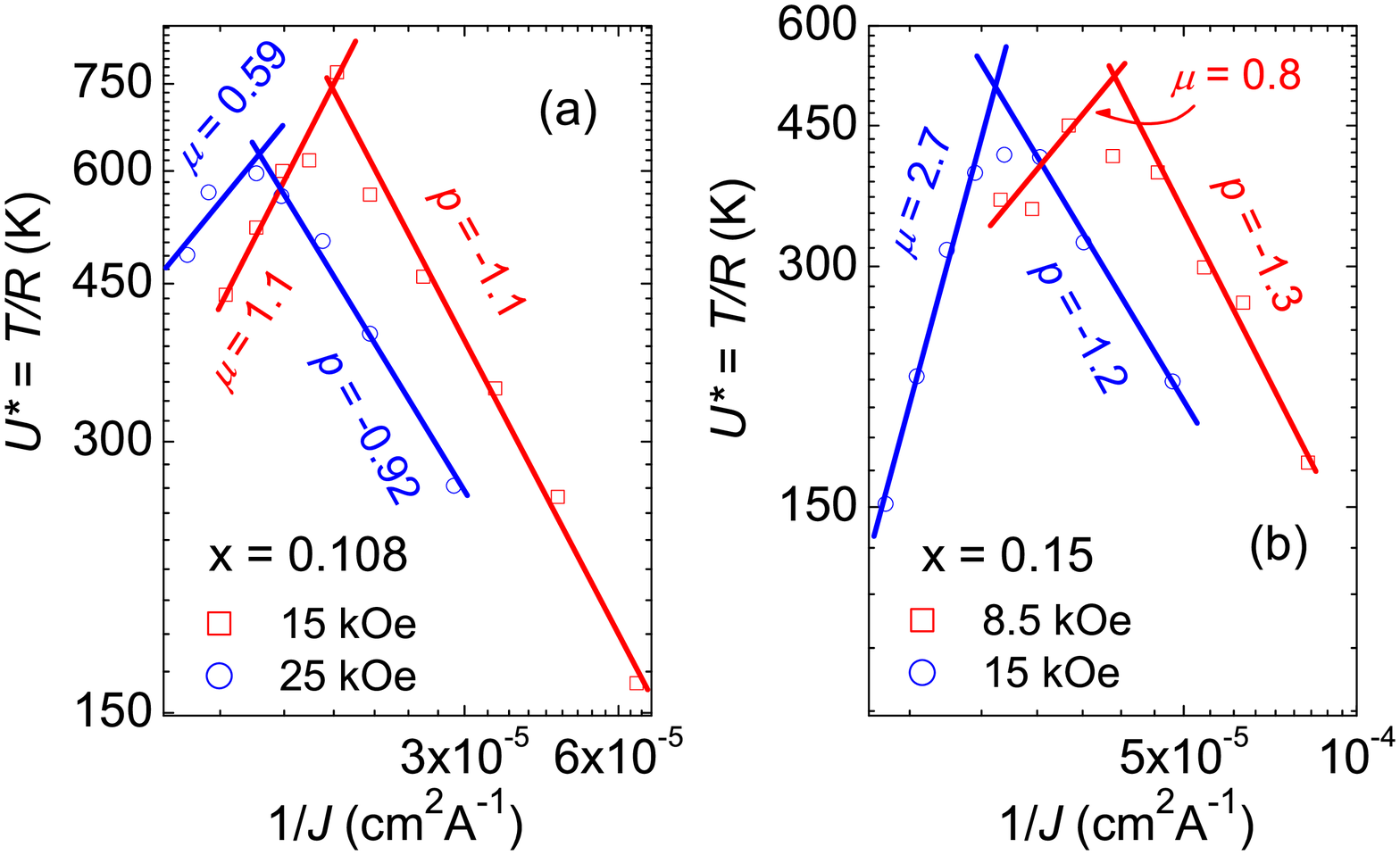}
\caption{\label{fig:Uvs.1/J} Activation energy ($U^* = T/R$) as a function of the inverse current density ($1/J$) for BaFe$_{2-x}$Ni$_x$As$_2$, x = (a) 0.108 and (b) 0.15 samples. For both samples, value of the parameters ($\mu$ and $p$) in each curve suggests the elastic to plastic pinning crossover and the crossover point is well matched with the $H_p$.}
\end{figure}

We exploited the temperature dependence of the relaxation rate, $R(T)$, to obtain the activation energy ($U^* = T/R$), and plotted it in Fig. \ref{fig:Uvs.1/J}  as a function of the inverse current density, 1/$J$, where $J$ is obtained using the Bean's critical state model, as discussed later. A $U^* vs. 1/J$ plot has been extensively used to investigate the vortex dynamics in pnictide superconductors \cite{wei06, ss17a, ss17b, tos12}. To relate the activation energy ($U^*$) with the critical current density ($J_c$), we used an expression from the theory of collective flux creep \cite{fei89}, $U^* = U_0(J_c/J)^{\mu}$, where, $\mu$ and $J_c$ depend on the dimensionality and size of the flux bundles under consideration \cite{fei89}. Using this expression, the exponent $\mu$ may be obtained by double logarithmic plot of $U^*$ vs. $1/J_c$, which is shown in Fig. \ref{fig:Uvs.1/J} (a) and (b) for x = 0.108 and 0.15 samples respectively. For a 3-dimensional system, the predicted values of exponent $\mu$ are reported as 1/7, 3/2 and 7/9, for single-vortex, small-bundle, and large-bundle regimes, respectively \cite{fei89, gri97}. However, the obtained $\mu$ values for x = 0.108 are 1.1 and 0.59 for $H$=15 and 25 kOe  and  $\mu$  for x=0.15 are 0.8 and 2.7 for $H$ = 8.5 and 15 kOe respectively (see Fig. \ref{fig:Uvs.1/J} (a), (b)). These $\mu$ values are different than the predicted ones, as found in other studies on several superconductors \cite{miu12, pro08, wei06, said10, yue13, hab11}. Similarly, values of he exponent ($p$) at higher temperature side (low $J$) is also found to be different than the predicted value for plastic creep ($p$ = 0.5) \cite{abu96, ss17b}. Although, the observed exponents in Fig. \ref{fig:Uvs.1/J} (a,b) are different than the expected values, the plots of $U^*$ vs. $1/J$ in the present study suggest a crossover in the pinning mechanism is responsible for SMP. 

\begin{figure}
\centering
\includegraphics[height=10cm]{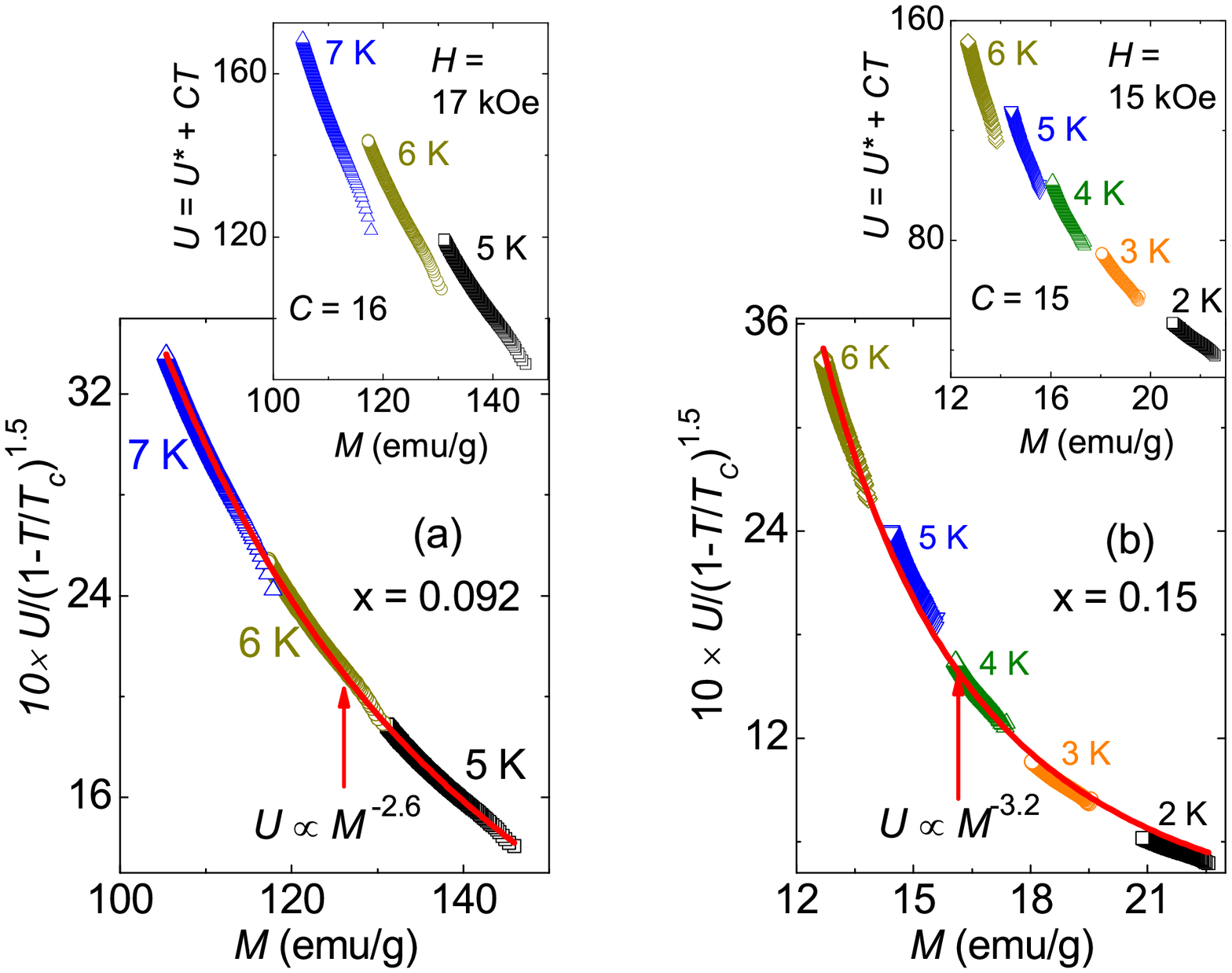}
\caption{\label{fig:Uvs.M} Variation of activation energy ($U$) scaled with a function, $g(T/T_c) = (1 - T/T_c)^{1.5}$, is plotted as a function of magnetization ($M$), for BaFe$_{2-x}$Ni$_x$As$_2$, x = (a) 0.092, and (b) 0.15 samples. Each scaled curve follow the  power law behaviour. Inset of each panel shows $U(M)$ before scaling by $g(T/T_c)$ function. Similar results are also observed for x = 0.108 sample (not shown here).}
\end{figure}

\begin{figure}
\centering
\includegraphics[height=12cm]{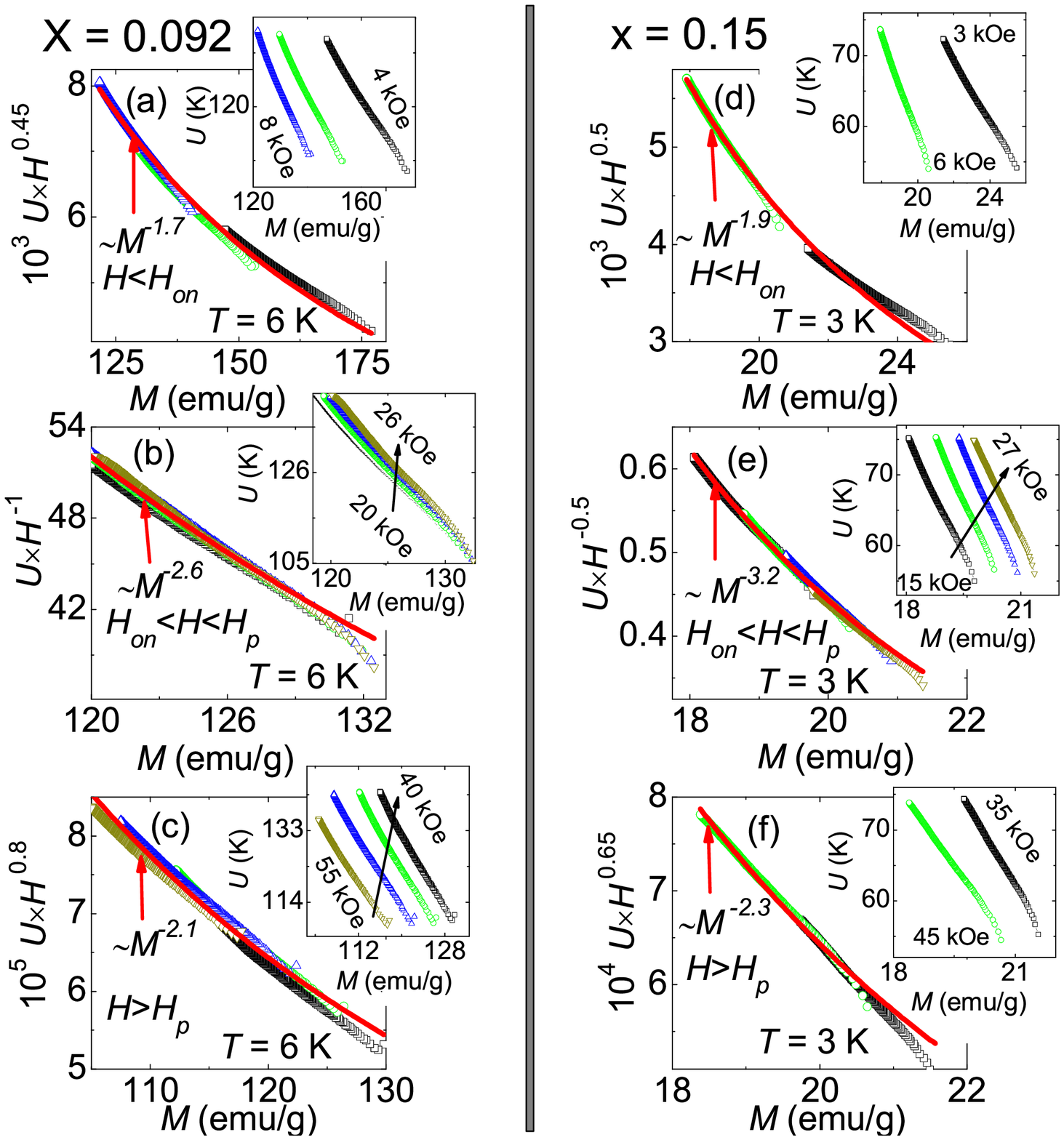}
\caption{\label{fig:EL-PL} Activation energy ($U$) scaled using the theory of collective flux creep is plotted as a function of magnetization ($M$) in panles, (a)-(c)for x = 0.092 and panels (d)-(f) for x = 0.15, in different magnetic field regimes (see text). Each sample shows the collective (elastic) to plastic creep crossover. Each inset shows the $U$ vs. $M$ without scaling in different magnetic field regimes. Similar results are also observed for x = 0.108 sample (not shown here).}
\end{figure}

In order to confirm a possible pinning crossover, observed in Fig. \ref{fig:RT_All} and Fig. \ref{fig:Uvs.1/J} for $R(T)$ and $U^* (1/J)$ curves respectively, we plotted the activation energy, $U$, of the, x = 0.092, and 0.15, samples, as a function of magnetization, $M$, by invoking the method developed by Maley $et$ $al$ \cite{mal90} (Similar results are observed for x = 0.108, but the not shown here). Such methodology has been widely used to investigate the vortex dynamics associated with SMP in different iron pnictide superconductors \cite{mal90, yon18, wei06, ss17a, said10, ahm17}, and is expressed below.
 
\begin{equation} \label{eq:1}
U = -Tln[dM(t)/dt] + CT,
\end{equation}

where $C$ is a constant which depends on the hoping distance of the vortex, the attempt frequency and the sample size. The activation energy as a function of magnetization is plotted in Fig. \ref{fig:Uvs.M} (a, b) for x = 0.0.092 and x = 0.15 samples respectively. The insets in Figs. \ref{fig:Uvs.M} (a, b) show the $U$ vs. $M$ curves for x = 0.0.092, and 0.15 samples using $C$ = 16, 25, 15 respectively. Similar values have been previously reported \cite{hen91,ss17b, said10}. The $U$ vs. $M$ curves for each sample, as shown in the insets figures, do not show a smooth behaviour. The curves showing a smooth power law behaviour are obtained after divided $U$ by $g(T/T_c)$ = $(1-T/T_c)^{1.5}$, as suggested in Ref. \cite{hen91} and verified in numerous studies \cite{ss17a, ss17b, said10, ahm17, yon18}. The smooth curves of $U/(1-T/T_c)^{1.5}$ vs. $M$ are shown in each main panel of Fig. \ref{fig:Uvs.M}. The values of the parameter $C$ used in Fig. \ref{fig:Uvs.M} are employed to extract the activation energy from the magnetic relaxation data in different field regimes, such as, $H$ $<$ $H_{on}$, $H_{on}$ $<$ $H$ $<$ $H_p$ and $H$ $>$ $H_p$, in order to investigate the SMP behaviour in x = 0.092, 0.108 and x = 0.15 samples (Results for x = 0.108 are not shown here).
 
To demonstrate the origin of the SMP in a series of BaFe$_{2-x}$Ni$_x$As$_2$, we plotted the activation energy as a function of magnetization, $U(M)$, shown in the inset of each panel of Fig. \ref{fig:EL-PL}. For x = 0.092, and 0.15 samples, the $U(M)$ curves were plotted for $T$ = 6 K, and 3 K respectively in three different magnetic field regimes. These $U(M)$ curves were analyzed in terms of the theory of collective flux creep \cite{fei89, abu96}, in which, the activation energy is defined as-

\begin{equation} \label{eq:2}
U(B,J) = B^{\nu}J^{-\mu} \approx H^{\nu}M^{-\mu},
\end{equation}
where, the exponents $\nu$ and $\mu$ depend on the specific pinning regime. According to the collective creep theory, the activation energy ($U$) increases with the magnetic field ($H$) and if the activation energy decreases with increasing magnetic field, it is suggestive of plastic creep behaviour \cite{abu96}. Therefore, in equation \ref{eq:2}, the positive value of $\nu$ suggests a collective creep mechanism and similarly, a negative value indicates plastic creep \cite{fei89, abu96}. In order to examine the collective (elastic) to plastic creep crossover in the samples, we scaled $U$ with $H^{\nu}$ for each sample under investigation in different magnetic field regimes for different values of $\nu$, as shown in Fig. \ref{fig:EL-PL}. The positive and negative values of the exponent $\nu$ in $H_{on}$ $<$ $H$ $<$ $H_p$ and $H$ $>$ $H_p$ magnetic field regions for each sample (x = 0.092 and 0.15), clearly demonstrate the collective (elastic) to plastic creep crossover as the origin of SMP in BaFe$_{2-x}$Ni$_x$As$_2$. Similar, elastic to plastic creep crossover is also observed for x = 0.108 sample, as the origin for SMP, but results are not shown here. In Figs. \ref{fig:EL-PL} (a), and \ref{fig:EL-PL} (d), the scaling of $U$ vs. $M$ curves for $H$ $<$ $H_{on}$ also shows the negative value of $\nu$ which would indicate the unphysical plastic creep nature. However, such behaviour is observed in other studies and has been well explained in terms of single vortex pinning (SVP) \cite{abu96, ss17a, ss17b, said10, wei06}. The crossover from SVP to collective creep renders a peak at $H_{on}$, which is entirely different in nature than the SMP at $H_p$.

\subsection{Critical current density and pinning behaviour} 

\begin{figure}
\centering
\includegraphics[height=10cm]{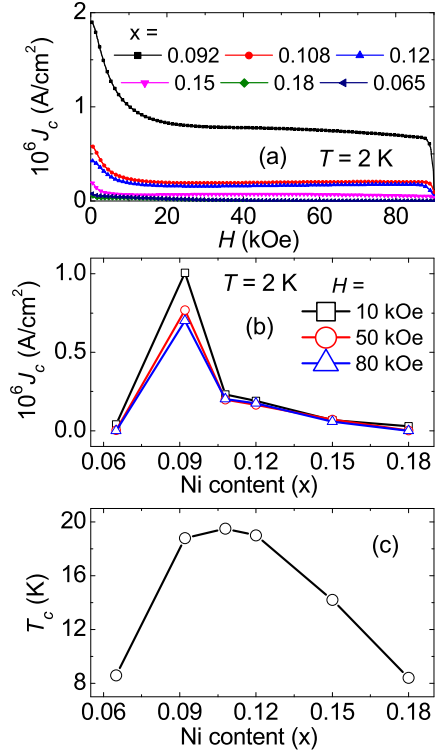}
\caption{\label{fig:Jcvsx} (a) Comparison of the magnetic field dependence of the critical current density, $Jc(H)$, at $T$ = 2 K, between distinct doping (x) contents in BaFe$_{2-x}$Ni$_x$As$_2$ superconductors. (b) Critical current density at $T$ = 2 K plotted as a function of Ni content (x). The maximum in $J_c$(x) corresponds to x = 0.092. (c) Doping dependence (x) of the superconducting transition temperature ($T_c$). Peak in $T_c$(x) corresponds to the x = 0.108, however, the optimal doping for BaFe$_{2-x}$Ni$_x$As$_2$ superconductors is x = 0.1 \cite{yan11}.}
\end{figure}

The magnetic field dependence of critical current density, $J_c(x)$, at $T$ = 2 K, for each Ni content is shown in Fig. \ref{fig:Jcvsx} (a). Bean's critical state model \cite{cha64} is exploited to extract the $J_c$, using, $J_c$ (A cm$^{-2}$) = 20$\Delta M$/$a(1-a/3b)$, where, $\Delta M$ (emu cm$^{-3}$) is the difference between the upper and lower branches of the isothermal $M(H)$ curves and $a$, $b$ are the dimensions of a rectangular shaped sample ($a<b$ in cm) perpendicular to the applied magnetic field direction \cite{ss17b, son16}. The critical current density ($J_c$) in iron-based superconductors is quite important for their potential use in technological applications \cite{ila15, son16}. The maximum, $J_c$ $\approx$ 2 MA/cm$^2$, is observed for the slightly underdoped sample, x = 0.092, in the zero field limit at $T$ = 2 K. On the other hand, for overdoped compounds (x = 0.108, 0.12, 0.15), $J_c$ is found to be higher than the threshold limit for technological application ($\approx$ 10$^5$ A/cm$^2$) in the zero magnetic field limit, even at liquid helium temperature ($T$ = 4.2 K). For further Ni doping, x = 0.18, $J_c$ decreases to the 10$^4$ A/cm$^2$ order of magnitude. It is found that the $J_c$ corresponding to the optimal doping (x = 0.1) \cite{li11, sha13} is smaller than the slightly underdoped (x = 0.9) regime \cite{sun09}, as has been also observed in the case of Ba$_{1-x}$K$_x$Fe$_2$As$_2$ \cite{son16}. The $J_c$ values above 10$^5$ A/cm$^2$ in overdoped samples and even more than 1MA/cm$^2$ in slightly underdoped compound makes BaFe$_{2-x}$Ni$_x$As$_2$ a potential candidate for application purpose \cite{ila15}. 

Figure \ref{fig:Jcvsx} (b) shows the behavior of $J_c$ as a function of Ni content (x) measured at $T$ = 2 K for different magnetic field values. It is interesting to see that $J_c(x)$ shows a spike like behaviour at x = 0.092 for each curve plotted for $H$ = 10 kOe, 50 kOe, 80 kOe. Interestingly, the $J_c(x)$ curve shown in Fig. \ref{fig:Jcvsx} (b) is distinctively different than the $T_c(x)$ plot presented in Fig. \ref{fig:Jcvsx} (c) which shows a broad dome like behaviour \cite{yan11, wang15} instead the spike like peak shown in Fig. \ref{fig:Jcvsx}. Such behaviour between $J_c(x)$ and $T_c(x)$ was also seen by Song $et$ $al$, in Ba$_{1-x}$K$_x$Fe$_2$As$_2$ \cite{son16}.

\begin{figure}
\centering
\includegraphics[height=10cm]{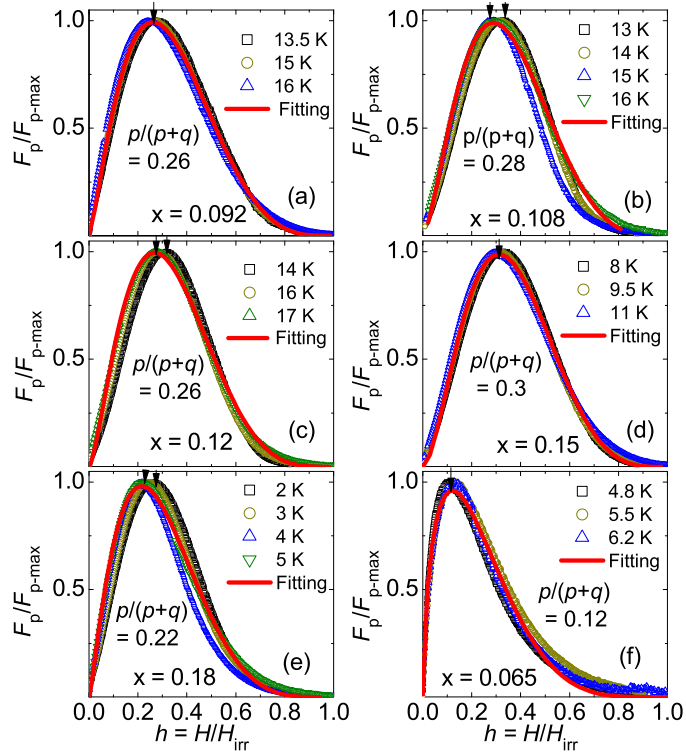}
\caption{\label{fig:Fp_All} The normalized pinning force density ($F_p/F_{p-max}$) as a function of reduced magnetic field, $h$ = $H/H_{irr}$ for for x = (a) 0.092, (b) 0.108, (c) 0.12, (d) 0.15, (e) 0.18, and (f) 0.065 samples. For each sample, data collected at different temperatures scaled to a single curve and the solid line is fit to the scaled curve using, $f_p = A(h)^p (1-h)^q$, where, the parameters $p$ and $q$ defines the pinning characteristics of the sample.}
\end{figure}

To investigate the pinning behaviour in BaFe$_{2-x}$Ni$_x$As$_2$, we estimate the pinning force density using, $Fp = J_c \times H$, where, $J_c$ is the critical current density and $H$ is the magnetic field. The normalized pinning force density is plotted as a function of reduced magnetic field ($h$ = $H$/$H_{irr}$) in Fig \ref{fig:Fp_All} (a-f) and is analyzed using the model developed by Dew-Hughes \cite{dew74}, which has been widely used in many other studies \cite{kob98, wei06, sha13, mat13}. The magnetic irreversibility field ($H_{irr}$) is extracted by considering the magnetic field value where $J_c$ $\le$ 50 A/cm$^2$, below which the $J_c$ decreases to the noise level. The scaling of the normalized pinning force curves shows a single peak for each sample under study. However, a close inspection of the scaled curves for different $T$ show  a slightly poor scaling for x = 0.108, 0.12 and 0.18 samples, with two nearby peaks, as shown with arrows in Fig. \ref{fig:Fp_All} (b), (c) and (e). On the other hand samples with  x = 0.092, 0.15 and 0.065 shows a good scaling with only one peak (see Fig. \ref{fig:Fp_All} (a), (d) and (f)). A peak behaviour of the scaled curve of pinning force suggests a single dominating pinning behaviour, which may be described in terms of a mathematical expression, $F_p/F_{p-max} = A (h)^p (1-h)^q$, where, $A$ is a multiplicative factor, $F_{p-max}$ is the maximum pinning force density at constant temperature, the parameters $p$ and $q$ provide the details about the pinning mechanism and the peak position is defined by $p/(p+q)$ \cite{kob98, kob16}. This expression was used to fit the scaled pinning force data shown in Fig. \ref{fig:Fp_All} (a-f), where, the solid line represents the fitting. The obtained parameters $A$, $p$, $q$ and the peak position $p/(p+q)$ for each sample are presented in the table \ref{table:parameters}.    

\begin{table}[ht]
\caption{Parameters obtained by fitting the expression, $F_p/F_{p-max} = A (h)^p (1-h)^q$, to the experimental curves $F_p/F_{p-max}$ vs. $h$} 
\centering 
\begin{tabular} {c c c c c} 
\hline\hline 
Samples & $A$ & $p$ & $q$ & $p/(p+q)$\\ [0.5ex] 

\hline 
x = 0.092 & 17.3 & 1.3 & 3.6 & 0.26 \\ 
x = 0.108 & 25.4 & 1.5 & 3.8 & 0.28 \\
x = 0.12 & 24.5 & 1.4 & 4.0 & 0.26 \\
x = 0.15 & 35.1 & 1.7 & 3.9 & 0.3 \\
x = 0.18 & 18.2 & 1.2 & 4.4 & 0.22 \\
x = 0.065 & 6.4 & 0.6 & 4.5 & 0.12 \\ [1ex]
\hline 
\end{tabular}
\label{table:parameters} 
\end{table}

It is known from the Dew-Hughes model that the high value of the peak position ($h >$ 0.33)  is an indication of dominant $\delta T_c$ pinning and peak position lower than, $h$ = 0.33, suggests the dominant role of $\delta l$ pinning and point like pinning centers\cite{dew74, kob98, kob16, per13}. Therefore, the peak positions shown in table \ref{table:parameters} indicates the $\delta l$ pinning behaviour for almost all investigated samples. However, for x = 0.065 (highly underdoped), the peak position is found to be 0.12, which is quite smaller than the overdoped and nearly optimaly doped samples. This scenario suggests that the pinning behaviour in overdoped and underdoped regimes are quite different in nature. It is to be noted that the peak position of the scaled pinning force curves is found at $h$ $\sim$ 0.3 in the study on a slightly underdoped, x = 0.09 sample \cite{sun09}. In other studies of optimally doped samples (x = 0.1), the scaled pining force curves show $h \ge$ 0.4 \cite{su14, sha13}, which would suggest the $\delta T_c$ pinning. However, Ref. \cite{su14} suggests the dominance role of $\delta T_c$ pinning in the sample, whereas, Ref. \cite{sha13} claims a strong signature of $\delta l$-type pinning. As we know the peak position in the scaled pinning force curves is dependent on the value of irreversibility field, $H_{irr}$, and based on the criterion to chose the $H_{irr}$, one may get a smaller or larger value of peak position ($h$). This shows that a method based on the peak position to describe the pinning mechanism is not robust enough. Therefore, we used the another approach to clear the ambiguity between the $\delta l$ and $\delta T_c$-type pinning in Ni-doped 122 superconductors. 

\begin{figure}
\centering
\includegraphics[height=10cm]{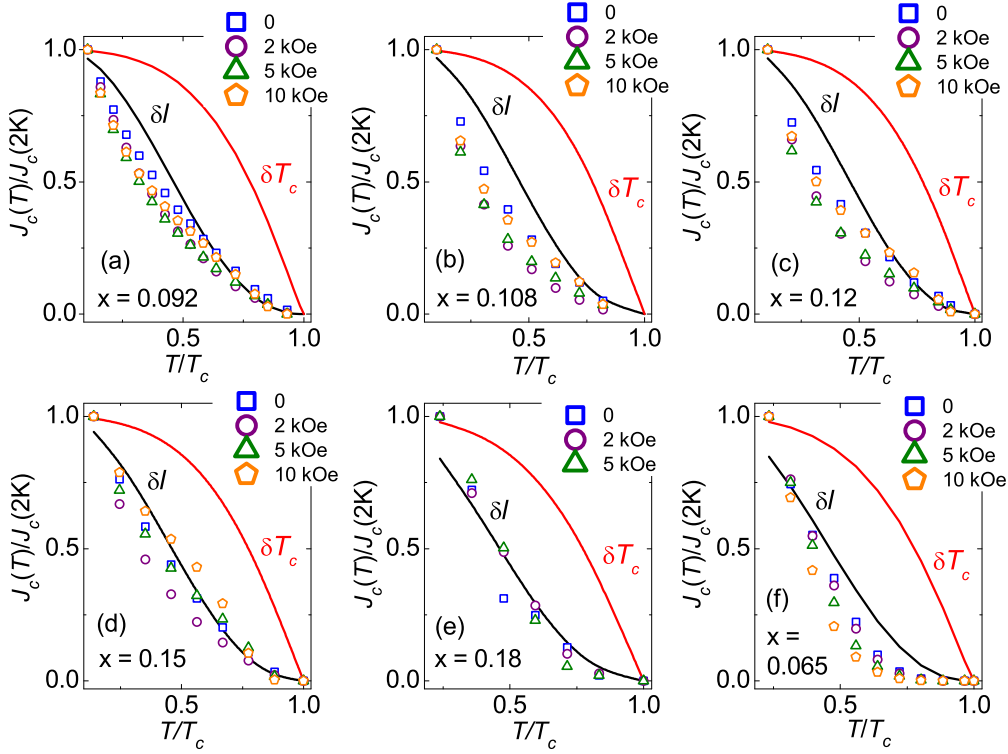}
\caption{\label{fig:DL_DTc} The normalized critical current density, $J_c/J_c(0)$, as a function of reduced temperature, $T/T_c$} for x = (a) 0.092, (b) 0.108, (c) 0.12, (d) 0.15, (e) 0.18, and (f) 0.065 samples. The solid lines present the $\delta l$ and $\delta T_c$ pinning models. Each sample shows close resemblance with the $\delta l$-type pinning mechanism.
\end{figure}

In order to explore the nature of pinning in a series of BaFe$_{2-x}$Ni$_x$As$_2$ superconductors, we investigate the temperature dependence of $J_c$ at different magnetic fields and used the model developed by Griessen $et$ $al$ \cite{gri94}. In this model, the pinning due to the spatial variation of the charge carrier mean free path, $\delta l$, and the spatial variation of the superconducting transition temperature, $\delta T_c$, have been described as, $\delta l$ ($J_c(t)/J_c(0)=(1-t^2)^{5/2}(1+t^2)^{-1/2}$) and $\delta T_c$ ($J_c(t)/J_c(0)=(1-t^2)^{7/6}(1+t)^{5/6}$). This model has been widely accepted to investigate the nature of vortex-pinning in superconductors \cite{vla15, gho10}. Figure \ref{fig:DL_DTc} (a-f), shows $J_c(T)/J_c(2 K)$ vs. $T/T_c$ plots at different constant $H$ and suggest the close resemblance with the $\delta l$-type pinning mechanism in all six samples under study. This result is consistent with the observation by Shahbazi $et$ $al$ \cite{sha13}. Bitter decoration patterns on optimally doped and overdoped, BaFe$_{2-x}$Ni$_x$As$_2$, show a highly inhomogeneous including large and small-scale stripe-like vortex patterns \cite{li11} preferably due to the dominant role of $\delta l$-type pinning. 

\section{Summary and Conclusion}

In summary, we studied a series of high quality BaFe$_{2-x}$Ni$_x$As$_2$, pnictide superconductors to investigate the doping dependence of the SMP, critical current density and the pinning characteristics. The SMP feature is observed in all samples except in a highly underdoped one, x = 0.065. Interestingly, for x = 0.092, the SMP feature is not prominent at low temperatures but is clearly visible above $T$ = 5 K. Temperature dependence of the relaxation rate, $R(T)$, suggest a pinning crossover, whereas, it's magnetic field dependence, $R(H)$, at different isothermals shows a peak structure. The peak position in $R(H)$, $H_{sp}$ lies in between the characteristic fields $H_p$ and $H_{on}$ associated with the SMP. In reference with other studies, such behaviour is described in terms of vortex-lattice structural phase-transition, which is followed by a pinning crossover. In order to confirm the pinning crossover, magnetic relaxation data was used to extract the activation energy ($U$) and was analysed using Maley’s method and collective pinning theory. The analysis unambiguously shows the collective (elastic) to plastic creep crossover as the origin of the SMP in Ni-doped BaFe$_2$As$_2$ superconductors. Such pinning crossover may be accompanied with a vortex-lattice structural phase transition below $H_p$. The critical current density ($J_c$) estimated using the Bean's critical state model is found to be larger than the threshold limit ($>$ 10$^5$ A/cm$^2$) considered for the technological relevance and even exceeds 1MA/cm$^2$ for x = 0.092 sample at low temperatures. However, for highly overdoped (x = 0.18) and underdoped (x = 0.065), the observed $J_c$ is lower than the threshold limit. The pinning behaviour in the samples is analyzed by plotting the normalized pinning force density ($F_p/F_{p-max}$) as a function of reduced magnetic field ($h$ = $H$/$H_{irr}$), which suggests the point like pinning centers in the samples. The plot of reduced temperature ($T/T_c$) dependence of the normalized critical current density ($J_c(T)/J_c(2 K)$) suggests that the pinning in the sample is related to the variation of the charge carrier mean free path ($\delta l$-type pinning).  

\section*{Acknowledgements}
SS acknowledges a post-doctoral fellowship from FAPERJ (Rio de Janeiro, Brazil), processo: E-26/202.848/2016. SSS and LG are supported by CNPq and FAPERJ. LFC is funded by The Leverhulme Trust grant number RPG-2016-306. The work at IOP, CAS is supported by the National Natural Science Foundation of China (11374011 and 11674372), the Strategic Priority Research Program (B) of the Chinese Academy of Sciences (CAS) (XDB07020300 and XDB25000000), and the Youth Innovation Promotion Association of CAS (2016004).

\section*{References} 



\end{document}